\documentstyle[mbf,epsf]{ptptex}

\title{Fluctuation modes in color-superconductors
}
 
\author{Masaharu {\sc Iwasaki}, Keiji {\sc Yamaguchi}$^a$ and Osamu {\sc Miyamura}$^a$
}
 
\inst{Department of Physics, Kochi University, Kochi 780-8520
\\
$^a$ Department of Physics, Hiroshima University, Higashi-Hiroshima 739-3526
}
 
\recdate{
}

\abst{
We investigate fluctuation effects of a gap parameter in color-superconductors. The fluctuation modes in the super phase are described by two scalar fields of diquarks. One of them is a Nambu-Goldstone boson and the other is a diquark boson whose mass is about twice of the gap energy (an extended quasi-supersymmetry). In the normal phase the fluctuation becomes a precursory (soft) mode whose amplitude increases near the critical temperature.
}

\begin{document}
\maketitle

\section{Introduction}
In many-body systems, spontaneous symmetry breaking \cite{NJL61} is a key concept in understanding the structure of the vacuum state. For example the gauge invariance is broken in the ground state of the superconductors and is restored at higher temperature than the critical point, $T_c$. The matter is in the Nambu-Goldstone (NG) phase at $T<T_c$ and in the Wigner phase at $T>T_c$. There is an order (gap) parameter that discriminates between such two phases; its value is positive in the NG phase and zero in the Wigner one. Such an order parameter is connected to a mean field in the framework of Hartree-Fock (Bogoliubov) approximation.

Usually the order parameter is fluctuating around the mean field. This fluctuation is also seen in the Wigner phase and increases near the critical temperature. The large fluctuation in the Wigner phase is a precursory mode of the spontaneous symmetry breaking and is called a soft mode. There are several examples of the soft modes, such as the paramagnon in the ferromagnet, the pairing vibration in spherical nuclei and the paracurrent in the superconductors. In hadron physics, Hatsuda and Kunihiro \cite{HK84}-\cite{HK85-2} found similar phenomenon concerning the chiral symmetry. The order parameter of this symmetry is $\langle q\bar{q}\rangle$ where $q$ means a quark field and its fluctuation modes in the NG phase are pions and $\sigma$ mesons. The precursor of this symmetry is discussed by them in \cite{HK85-2}.

Recently another symmetry breaking has been interested in hadron physics; the order parameter is $\langle qq\rangle $ and is called color-superconductivity \cite{BL84}-\cite{W00}. It is the purpose of this paper to discuss the soft modes concerning this order parameter. Particularly its precursor will be investigated near the critical temperature. It is also pointed out that this soft mode contributes to the cooling of the hot quark matter.

\section{Effective action}
We consider the quark matter with three flavors. Its Lagrangian is supposed to be given by the following four-Fermi interaction,
\begin{equation}
{\mathcal L}=\bar{\psi}(i\gamma\partial+\mu\gamma^{0})\psi+g\sum_{a}(\bar{\psi}\gamma_{\mu}\lambda^{a}\psi)(\bar{\psi}\gamma^{\mu}\lambda^{a}\psi), 
\end{equation}
where $g$, $\mu$ and $\lambda^a$ denote the coupling constant, chemical potential and flavor SU(3) matrix, respectively. By using Fierz transformation on the right-hand side, we have
\begin{equation}
{\mathcal L}=\bar{\psi}(i\gamma\partial+\mu\gamma^0)\psi+\frac{2}{3}g{\sum_{a,b=2,5,7}}'(\bar{\psi}\gamma_{5}C\lambda^{a}\Lambda^{b}\bar{\psi}^{t})(\psi^{t}C^{-1}\gamma^{5}\lambda^{a}\Lambda^{b}\psi).
\end{equation}
Here we have left only the most attractive terms, which are spin-singlet, color-triplet and flavor-triplet pairs. This means that the $\lambda_{a}$ and $\Lambda_{b}$ are restricted to antisymmetric SU(3) matrices. 

In order to discuss the system at finite temperature, let us introduce a partition function represented in the functional form,
\begin{equation}
Z=\int {\mathcal D}\bar{\psi} {\mathcal D}\psi\exp (-S),
\end{equation}
where $S$ is a Euclid action defined by $S=\int d\tau d{\vec r}{\mathcal L}$ ($\tau=it$). 

Now we introduce a scalar field $\phi_{\rho}(x)$ representing a wave function of a pair of quarks. Then the partition function can be rewritten as
\begin{equation}
Z=\int {\mathcal D}\bar{\psi}{\mathcal D}\psi{\mathcal D}\phi^{*}{\mathcal D}\phi \exp (-S'),
\end{equation}
where the corresponding Lagrangian is given by
\begin{equation}
{\mathcal L'}=\bar{\psi}(i\gamma\partial+\mu\gamma^0)\psi - \frac{2}{3}g\sum_{\rho}(\bar{\psi}\gamma_{5}C T^{\rho}\bar{\psi}^{t}\phi{\rho} + h.c.)+|\phi_{\rho}|^2.
\end{equation}
The $T^{\rho}$ denotes $\lambda_{a}$ and $\Lambda_{b}$. To calculate this integral, it is convenient to use Fourier transformations,
\begin{eqnarray}
\psi(x)&=&\sum_{p,s}\psi(p,s)u(\vec{p},s) e^{i(\vec{p}\vec{x}-\omega\tau)},\\
\phi(x)&=&\sum_{q}\phi(q) e^{i(\vec{q}\vec{x}-\nu\tau)},
\end{eqnarray}
where the Matsubara frequencies are defined by $\omega=(2n+1)\pi T$ and $\nu=2m\pi T$. Substituting these equations into (5), the action is written in the Nambu representation as follows \cite{N60}:
\begin{eqnarray}
S'&=&\frac{\beta}{2}\sum_{p,p',s}
\left(\begin{array}{c}\psi^{*}(p,s) \\ s\psi(\tilde{p}',s) \end{array}\right)^{t} \left(\begin{array}{cc}i\omega-\xi_{p} & \Delta(p,\tilde{p}') \\ \Delta^{*}(\tilde{p}',p) & i\omega'+\xi_{p'} \end{array}\right) \left(\begin{array}{c}\psi(p,s) \\ s\psi^{*}(\tilde{p}',s) \end{array}\right) \nonumber \\
& &- \beta\sum_{q}|\psi_{\rho}(q)|^2.
\end{eqnarray}
Here the uses are made of $p=({\vec p},\omega)$ and $q=({\vec q},\nu)$. The gap parameter on the right-hand side is defined by
\begin{equation}
\Delta(p,\tilde{p}')\equiv \sqrt{\frac{8g}{3}}u^{\dag}(\vec{p},s)u(\vec{p}',s)T^{\rho}\phi_{\rho}(p'-p).
\end{equation}
If we carry out the integration with respect to the quark fields, we obtain the partition function expressed in terms of scalar (pairing) fields:
\begin{equation}
Z=\int {\mathcal D}\phi^{*}{\mathcal D}\phi \exp[-S_{\rm eff}],
\end{equation}
where the effective action $S_{\rm eff}$ is defined by
\begin{equation}
S_{\rm eff}=-{\rm Tr}\ln (\beta G^{-1})-\beta\sum_{q}|\phi_{\rho}(q)|^2.
\end{equation}
The effective action is described only by the pairing field and will play an important role in our later discussions.

\section{Fluctuations}
To begin with we introduce a mean field approximation for the effective action. The path integral of the effective action is replaced by a mean field, which should be stationary field in the path integral. We assume the following constant field,
\begin{equation}
\Delta_{0}(p,\tilde{p}')\equiv \delta_{p,p'}\sqrt{\frac{8g}{3}}\phi_{0}{\sum_{a}}'\lambda_{a}\otimes \Lambda_{a}.
\end{equation}
This pair has the most symmetric state in color-flavor space and known as the color flavor locking state \cite{ARW99}. The constant $\phi_0\equiv \phi(0)$ will be determined later.

Next let us consider fluctuation around the constant field. Provided we write the inverse propagator as
\begin{eqnarray}
G^{-1}&=&\left(\begin{array}{cc}i\omega-\xi_{p} & \Delta(p,\tilde{p}') \\ \Delta^{*}(\tilde{p}',p) & i\omega'+\xi_{p'} \end{array}\right) + \sqrt{\frac{8g}{3}}\left(\begin{array}{cc} 0 & \tilde{\phi}(p,\tilde{p}')\hat{\epsilon} \\ \tilde{\phi}^{*}(\tilde{p}',p)\hat{\epsilon} & 0 \end{array}\right) \nonumber \\
&\equiv& G_{0}^{-1}+\Delta G^{-1},
\end{eqnarray}
the fluctuation is described by the second term on the right-hand side. Substituting this equation into the effective action, we get
\begin{equation}
S_{\rm eff}=S^{(0)}_{\rm eff}+S^{(1)}_{\rm eff}+S^{(2)}_{\rm eff}\cdot \cdot \cdot.
\end{equation}
The first term is nothing but the mean field (BCS) action. The second term should vanish if the mean field is chosen to be stationary one; this condition is proved to be equivalent to the gap equation. The third term describes the fluctuation and is given by
\begin{equation}
S_{eff}^{(2)}=\frac{1}{2}{\rm Tr}(G_{0}\Delta G^{-1})^2 +\beta\sum_{q}|\tilde{\phi}(q)|^2.
\end{equation}
The BCS propagator on the right-hand side is written as
\begin{equation}
G_{0}=\left(\begin{array}{cc} -i\omega-\xi_{p} & \Delta(p,\tilde{p}') \\ \Delta^{*}(\tilde{p}',p) & -i\omega'+\xi_{p'} \end{array}\right) (\omega_{n}^2 +\hat{E}_{p}^2)^{-1}.
\end{equation}
The $\hat{E}_p$ denotes a quasiparticle energy and defined by $\hat{E}^2_{p}=\xi^2_{p}+\Delta^{\dag}_{0}\Delta_{0}$. Substituting this equation into (15) and rewrite it in the matrix form, the action of second order is transformed into
\begin{equation}
S_{\rm eff}^{(2)}=\sum_{q}
\left(\begin{array}{cc}\phi^{\dag}(q) & \phi(-q) \end{array}\right) \left(\begin{array}{cc} X_{q} & Y_{q} \\ Y^{*}_{q} & X^{*}_{q} \end{array}\right) \left(\begin{array}{c}\phi(q) \\ \phi^{*}(-q) \end{array}\right).
\end{equation}
The matrix elements on the right-hand side are defined by
\begin{eqnarray}
X_q&=&\frac{4g}{3}\sum_{p}{\rm tr}\left[\frac{(i\omega_p+\xi_p)(i\omega_{p-q}-\xi_{p-q})\hat{\epsilon}^2}{(\omega_{p}^2+\hat{E}_{p}^2)(\omega_{p-q}^2+\hat{E}_{p-q}^2)}\right]|N_{p,p-q}|^2+\frac{1}{2}\beta, \\
Y_q&=&\frac{4g}{3}\sum_{p}{\rm tr}\left[\frac{\Delta_{0}^{\dag}\Delta_{0}\hat{\epsilon}^2}{(\omega_{p}^2+\hat{E}_{p}^2)(\omega_{p-q}^2+\hat{E}_{p-q}^2)}\right]|N_{p,p-q}|^2,
\end{eqnarray}
where the ${\rm tr}$ denotes for a trace of color-flavor space.

Here we introduce two real scalar fields instead of the complex field $\phi(q)$,\begin{equation}
\sigma(q)\equiv \phi(q)+\phi^{*}(-q), \quad \pi(q)\equiv -i(\phi(q)-\phi^{*}(q)),\end{equation}
with the relations of $\sigma^{*}(q)=\sigma(-q)$ and $\pi^{*}(q)=\pi(-q)$. These fields are analogous to the $\sigma$ and $\pi$ fields in the chiral dynamics respectively. By using new fields, we get
\begin{equation}
S_{\rm eff}^{(2)}=\sum_{q}\left(\sigma^{*}(q)D_{\sigma}^{-1}(q)\sigma(q) + \pi^{*}(q)D_{\pi}^{-1}(q)\pi(q)\right),
\end{equation}
where two propagators are defined by $D_{\sigma}^{-1}(q)\equiv (X_q+X_{q}^{*}+2Y_q)/4$ and $D_{\pi}^{-1}(q)\equiv (X_q+X_q^{*}-2Y_q)/4$. It is evident that the sigma mode represents the amplitude mode of the gap parameter and the pi mode corresponds to the phase mode. In chiral dynamics, the former corresponds to a sigma meson and the latter does a pi meson respectively \cite{HK85}.

\section{Soft modes}
Now we discuss the fluctuation modes in detail. Our interest will be put on the fluctuation near the critical temperature. It is convenient that our discussions are separated into two cases:

(i) $\phi_0\neq 0$ ($T<T_c$):
Each mode corresponds to a field of the diquark with fractional baryon number. This particle is a boson but similar to the anti-quark concerning the color quantum number. The diquark masses should be calculated from zeros of the inverse propagators. For the pi mode, it is represented by
\begin{equation}
D^{-1}_{\pi}({\vec q}=0,\nu)=\frac{8g\beta}{3}\sum_{\vec p}{\rm tr}\left[{\hat \epsilon}^2\frac{(i\nu)^2}{(i\nu)^2-4{\hat E}_{p}^2}\frac{1-2f({\hat E}_p)}{2{\hat E}_p} \right],
\end{equation}
where the use is made of the gap equation. This equation means that the mass vanishes and the diquark is a Nambu-Goldstone boson. In the same way, the inverse propagator for the sigma mode is written as
\begin{equation}
D^{-1}_{\sigma}({\vec q}=0,\nu)=\frac{8g\beta}{3}\sum_{\vec p}{\rm tr}\left[
{\hat \epsilon}^2\frac{(i\nu)^2-4\Delta^2}{(i\nu)^2-4{\hat E}_{p}^2}\frac{1-2f({\hat E}_p)}{2{\hat E}_p} \right].
\end{equation}
It is found that if the $\Delta$ is a $\Delta_0\times {\rm 1}$, the mass of the $\sigma$ is equivalent to $2\Delta_0$. Noting that the effective quark mass is $\Delta_0$, we have the following mass relation: $M_{\pi}:M_{q}:M_{\sigma}=0:1:2$. This simple relation is called quasi-supersymmetry by Nambu \cite{MN89}. In our case, $\Delta$ has a complicated structure so that we have an extended quasi-supersymmetry relation. When the temperature approaches to the critical one, the $M_{\sigma}$ goes to zero. This implies that the fluctuation due to the $\sigma$ field is enhanced as expectedly.   

(ii) $\phi_0=0$ ($T>T_c$):
It is well known that even in the Wigner phase the fluctuation of order parameter increases near the critical temperature. This fluctuation is called soft mode or precursor. Noting that $X_q^{*}=X_{-q}$ and $Y_q=0$, the previous action (17) for the scalar field is described by
\begin{equation}
S_{\rm eff}^{(2)}=2\sum_{q}\phi^{\dag}(q)D^{-1}(q)\phi(q),
\end{equation}
where the inverse propagator is given by
\begin{equation}
D^{-1}(q)\equiv 2X_q=32g\sum_{p}\frac{|N_{p,p-q}|^2}{(i\omega-\xi_p)(i\omega-i\nu+\xi_{p-q})}+\beta.
\end{equation}
Carrying out the summation of the Matsubara frequencies, we get
\begin{equation}
D^{-1}(q)=32g\beta\sum_{\vec p}\frac{f(\xi_{p+\frac{1}{2}q})+f(\xi_{p-\frac{1}{2}q})-1}{\xi_{p+\frac{1}{2}q}+\xi_{p-\frac{1}{2}q}-i\nu}|N_{p,p-q}|^2 +\beta,
\end{equation}
where $f(x)\equiv (1+e^{\beta x})^{-1}$ is the Fermi distribution function.

To discuss behavior of the soft mode, let us investigate the spectral function of our system just as done in \cite{HK84}. First the retarded Green's function is obtained from (24) by analytic continuation, $i\nu\to \omega+i\epsilon$,
\begin{eqnarray}
D^{-1}_{R}(\omega)&=&\frac{2g\beta}{3\pi^2}\int_{-\infty}^{\infty}d\xi (\mu+\xi)^2 (2f(\xi)-1)\left(\frac{{\mathcal P}}{\xi-\frac{\omega}{2}}+i\pi\delta(\xi-\frac{\omega}{2})\right) +\beta,
\end{eqnarray}
where we have restricted ourselves to the ${\vec q}=0$ mode for simplicity. From this equation we get the spectral function,
\begin{equation}
S(\omega)\propto {\rm Im}D_{R}(\omega)={\rm Im}\frac{1}{D^{-1}_{R}(\omega)}. 
\end{equation}
Although it is not difficult to calculate the right-hand side of this equation, we can find the qualitative behavior of the spectral function for low frequency. First we notice that the imaginary part of the inverse propagator is proportional to $\omega$: ${\rm Im}(D^{-1}(\omega))\propto a\omega$. The proportional constant $a$ is a function of the temperature, $a(T)=a_{0}/T$. On the other hand, the real part ${\rm Re}(D^{-1}(\omega))$ is represented by $b(T)=b_{0}(T-T_c)$, because it vanishes at $T=T_c$. Therefore the spectral function is described by $S(\omega)\propto a\omega/(b^2+a^2 \omega^2)$. This function has a peak at $\omega=b/a$ and the height is $1/2ab$. As the temperature $T$ approaches to $T_c$, the position of the peak approaches to zero and the height increases. Thus we have shown that the spectral function is enhanced near the critical temperature.

Finally let us comment a characteristic phenomenon that this soft mode brings about. It was pointed out that the acceleration of cooling of quark stars is caused by the soft mode on chiral symmetry \cite{HK85-2}. Our soft mode is also expected to work in the same way. For example, two high-energy quarks in the hot quark star may be transformed into a soft diquark and cooled with emission of a pair of neutrinos via the following weak interaction process: $q_1+q_2\to q'_1+e^{-}+\bar{\nu_e}+q_2\to (q'_{1}q'_{2})_{\rm soft}+\bar{\nu_e}+\nu_e$. If Bose-Einstein condensation of the soft mode occurs, a huge number of neutrinos would be created so that very rapid cooling is realized in the quark stars just above the critical temperature. On the other hand, the cooling is strongly suppressed due to the color superconductivity below the same critical temperature as discussed by many authors \cite{R00}-\cite{ABR01}. Such a singular cooling behavior would be a possible sign for the color superconductivity in the hot quark matter.

\section*{Acknowledgements}

After almost completion of this work, we learned from Dr. T.Kunihiro that investigation of a similar nature is going on by his group. We would like to thank to Dr.T.Kunihiro for his discussions and encouragement. O.~Miyamura passed away during the completion of the paper. His coauthors would like to devote this work to his memory.

\end{document}